\DocumentMetadata{
 pdfversion=2.0,
 lang = en
 }
\documentclass[acmsmall,screen]{acmart}

\usepackage{amsmath,array,booktabs,float}
\usepackage{xurl}
\usepackage{epstopdf}
\hypersetup{hidelinks}
\acmJournal{TRETS}
\setcopyright{none}
\acmDOI{}

\begin{document}

\title{Exhaustive Evaluation of Fixed-Point Reciprocal Architectures for FPGAs}

\author{Navaneeth Kutty}
\email{navaneeth2310165@ssn.edu.in}
\orcid{0009-0007-9552-1034}
\affiliation{
  \department{Department of Electronics \& Communication Engineering}
  \institution{Sri Sivasubramaniya Nadar College of Engineering}
  \city{Chennai}
  \state{Tamil Nadu}
  \country{India}
}

\author{Anish Murali}
\email{anish2310158@ssn.edu.in}
\orcid{0009-0006-0900-3381}
\affiliation{
  \department{Department of Electronics \& Communication Engineering}
  \institution{Sri Sivasubramaniya Nadar College of Engineering}
  \city{Chennai}
  \state{Tamil Nadu}
  \country{India}
}

\author{Iacovos Ioannou}
\email{iakovos.ioannou@nup.ac.cy}
\affiliation{
  \department{Department of Computer Science}
  \institution{Neapolis University Pafos}
  \city{Pafos}
  \country{Cyprus}
}

\author{Priyan Mariappan}
\email{priyan2310182@ssn.edu.in}
\orcid{0009-0005-1938-3800}
\affiliation{
  \department{Department of Electronics \& Communication Engineering}
  \institution{Sri Sivasubramaniya Nadar College of Engineering}
  \city{Chennai}
  \state{Tamil Nadu}
  \country{India}
}

\author{Prabagarane Nagaradjane}
\email{prabagaranen@ssn.edu.in}
\affiliation{
  \department{Department of Electronics \& Communication Engineering}
  \institution{Sri Sivasubramaniya Nadar College of Engineering}
  \city{Chennai}
  \state{Tamil Nadu}
  \country{India}
}

\renewcommand{\shortauthors}{Kutty et al.}

\begin{abstract}
Fixed-point reciprocals are used in normalization and division on field-programmable gate arrays (FPGAs). Designs that share an output format can differ in worst-case error and in their need for memory and multipliers. The effect of table width, internal precision and refinement was evaluated for reciprocal computation in the Q16.16 and Q8.24 formats. Every representable input in the interval from 0.5 to 2.5 was tested against an exact integer reference. Five arithmetic paths were reconstructed from register-transfer-level source. Guarded interpolation, Newton-Raphson refinement and residual correction were then evaluated alongside polynomial and digit-recurrence baselines. Correct rounding was verified for all 131,073 Q16.16 and 33,554,433 Q8.24 inputs by adding one residual check to a guarded interpolated seed and one refinement step. An implementation with product reuse was verified at one input per cycle with a latency of 10 cycles in both formats. The Q8.24 implementation used nine digital signal processing slices in 7 series synthesis. The exact radix-2 control used no such slices and had a latency of 28 cycles. The reconstructed refined paths reduced the maximum relative error by factors of 205 to 223 when eight fractional bits were added. The evaluated designs provide reproducible choices for fixed accuracy budgets. In particular, correct rounding can be obtained on the tested interval by using a compact seed and checking the integer residual before the result is returned.
\end{abstract}




\maketitle

\raggedbottom
\setlength{\intextsep}{10pt plus 2pt minus 2pt}

\section{Introduction}\label{sec:intro}

A reciprocal unit is used when division by a changing positive input is needed in a fixed-point datapath on a field-programmable gate array (FPGA). Examples include vector normalization and the scaling of signal samples. A direct lookup table can avoid multiplications, but its error is set largely by the spacing of the table entries. Interpolation and multiplicative refinement can reduce the spacing error at the cost of additional arithmetic. Fixed-point quantization can then become the main source of error \cite{ErcegovacLang04}.

The study addresses two questions. The first is how much accuracy is gained by changing the format, table width and internal precision. The second is which designs meet the same accuracy requirement with fewer table bits or products. The input range, output format and exact reference are shared across all evaluated numerical methods. Hardware claims are tied to the available implementation evidence.

Five arithmetic paths were reconstructed from the available register-transfer-level (RTL) source \cite{KuttyArtifact}. Coefficient images and raw outputs for the final historical experiment were unavailable at the cited revision. Independent coefficients were generated for the reconstructions and the resulting outputs were scored on complete input grids. The effects of guard bits and residual correction were assessed under common rounding rules. Polynomial baselines were evaluated using published coefficients and a radix-2 recurrence was included as an exact control.

Related methods are reviewed in Section~\ref{sec:related}. The arithmetic is defined in Section~\ref{sec:model}. The input grids, controls and hardware procedures are described in Section~\ref{sec:evaluation}. Numerical and RTL results are presented in Section~\ref{sec:results}. Design choices and scope are discussed in Section~\ref{sec:discussion}, followed by the conclusions in Section~\ref{sec:conclusion}. Reproduction commands are given in Appendix~\ref{sec:reproduce}.

\section{Related Work}\label{sec:related}

Division architectures were classified by Oberman and Flynn \cite{Oberman97}. Table-based approximation and functional iteration for reciprocals are treated by Ercegovac and Lang \cite{ErcegovacLang04}. The convergence method introduced by Goldschmidt was developed for division by multiplicative correction \cite{Goldschmidt64}. The use of a piecewise-linear initial approximation for multiplicative division was examined by Ito et al. \cite{Ito97}. A fixed-point reciprocal based on a pipelined Newton-Raphson implementation without input normalization was presented by Libessart et al. \cite{Libessart17}. Fixed-point reciprocal and division architectures using digit recurrence and multiplicative iteration were compared on a Virtex-4 device by Kumar et al. \cite{Kumar17}.

A polynomial seed followed by Newton-Raphson refinement was used in the fixed-point divider developed by Rodr\'iguez-Garc\'ia et al. \cite{Rodriguez13}. In the multifunction design of Guti\'errez-Ramos et al., reciprocal computation was supported by a piecewise quadratic seed on a normalized interval \cite{Gutierrez26}. A polynomial correction of a power-of-two reciprocal linearization was proposed by Lunglmayr \cite{Lunglmayr21}. These two studies supply published coefficients for the numerical comparison. An additional exact control uses radix-2 non-restoring division, a method also described in the AMD Divider Generator product guide \cite{AMD_PG151}.

The comparison is organized around verified error over the same discrete input domain and with the same output precision. Published FPGA resource and timing results are tied to the implementations and devices used in those studies. A common numerical protocol is applied to the reference designs and alternatives (as described in Section~\ref{sec:evaluation}).

\section{Arithmetic Model}\label{sec:model}

Integer codes and scoring rules are defined first (as shown in Section~\ref{sec:codes}). The roles of table spacing and multiplicative refinement are then derived (as shown in Sections~\ref{sec:tables} and~\ref{sec:iteration}). Correct rounding is checked by an integer residual (as shown in Section~\ref{sec:residual}). Ideal-arithmetic seed widths are then compared at two relative-error targets (as shown in Section~\ref{sec:seed_planning}).

\subsection{Integer Codes and Error Measures}\label{sec:codes}

A signed 32-bit word in the Q$I.F$ format is used, where $I$ includes the sign bit and $F$ is the number of fractional bits. The notation $Q_F(\cdot)$ denotes rounding to the nearest multiple of $2^{-F}$ with ties toward the larger value. Let $S=2^F$, let $k$ be the positive input code and let $q$ be the output code. The represented quantities and exact rounded reference $q^\star$ are
\begin{equation}
x=\frac{k}{S},\qquad y=\frac{q}{S},\qquad
q^\star=\left\lfloor\frac{2S^2+k}{2k}\right\rfloor.
\label{eq:codes}
\end{equation}
The ideal reciprocal is $1/x$. The least significant bit (LSB) of the output represents $1/S$. The ideal absolute error $E_{\rm abs}$ in LSB, the ideal relative error $E_{\rm rel}$ in percent and the rounded-reference difference $E_{\rm code}$ in LSB are defined as
\begin{equation}
\begin{aligned}
E_{\rm abs}(k)&=\frac{|qk-S^2|}{k},\\
E_{\rm rel}(k)&=100\frac{|qk-S^2|}{S^2},\\
E_{\rm code}(k)&=|q-q^\star|.
\end{aligned}
\label{eq:errors}
\end{equation}
The residual in Equation~(\ref{eq:errors}) is computed using the integer codes from Equation~(\ref{eq:codes}). Maxima of absolute and relative error are evaluated independently because they can occur at different inputs. The ideal-reference root mean square error (RMSE) is evaluated as follows. For the tested set $\mathcal K$ of $M=|\mathcal K|$ codes, it is
\begin{equation}
E_{\rm rms}=\left(\frac{1}{M}\sum_{k\in\mathcal K} E_{\rm abs}(k)^2\right)^{1/2}.
\label{eq:rmse}
\end{equation}
Equation~(\ref{eq:rmse}) weights each tested code equally. The rounded-code RMSE and the mean ideal relative error are
\begin{equation}
\begin{aligned}
E_{\rm rms,code} &= \left(\frac{1}{M}\sum_{k\in\mathcal K}E_{\rm code}(k)^2\right)^{1/2},\\
\overline{E}_{\rm rel} &= \frac{1}{M}\sum_{k\in\mathcal K}E_{\rm rel}(k).
\end{aligned}
\label{eq:aggregate_error}
\end{equation}
The rounded-reference match count is the number of codes for which $E_{\rm code}(k)=0$. Exact rounding can still leave a nonzero ideal error.

\subsection{Table Seeds and Chord Bounds}\label{sec:tables}

For $a=0.5$ and $b=2.5$, the interval is divided into $2^N$ equal cells of width $\Delta=(b-a)/2^N$, with left boundaries $x_i=a+i\Delta$. The last cell includes $b$. A direct lookup returns the stored value at $x_i$. For exact coefficients and interior cells, the relative and absolute errors are
\begin{equation}
x/x_i-1=\frac{x-x_i}{x_i},\qquad
\frac{1}{x_i}-\frac{1}{x}=\frac{x-x_i}{x_i x}.
\label{eq:direct}
\end{equation}
On the input grid the largest interior offset is $\Delta-1/S$ when $\Delta S$ is an integer. The endpoint $b$ is assigned to the last cell and has offset $\Delta$. These two cases are evaluated separately in Equation~(\ref{eq:direct}). Quantizing a table entry can also give a negative signed error at a left boundary.

Let $\ell_i(x)$ be the chord through $1/x_i$ and $1/(x_i+\Delta)$. The ideal chord is formed as
\begin{equation}
\begin{aligned}
\ell_i(x)&=\frac{1}{x_i}+\frac{x-x_i}{\Delta}
 \left(\frac{1}{x_i+\Delta}-\frac{1}{x_i}\right),\\
x\ell_i(x)-1&=\frac{(x-x_i)(x_i+\Delta-x)}{x_i(x_i+\Delta)}.
\end{aligned}
\label{eq:chord}
\end{equation}
Equation~(\ref{eq:chord}) gives the relative bound $\Delta^2/[4a(a+\Delta)]$ in exact arithmetic. Division by $x$ converts the relative error to absolute error. An ideal reference scale of one output LSB corresponds to a relative fraction $x/S$ at input $x$. A sufficient global relative bound for a pointwise one-LSB guarantee is $a/S$. This distinction is used when seed widths are compared (as shown in Section~\ref{sec:tuning}). Coefficient and product errors are assessed separately through the integer models.

\subsection{Multiplicative Refinement}\label{sec:iteration}

For a current estimate $y_j$ and residual $r_j=1-xy_j$, a Newton-Raphson step in exact arithmetic is
\begin{equation}
y_{j+1}=y_j(2-xy_j),\qquad r_{j+1}=r_j^2 .
\label{eq:nr}
\end{equation}
Convergence of Equation~(\ref{eq:nr}) is governed by $|r_0|<1$. After $j$ exact steps the residual magnitude is $|r_0|^{2^j}$. In the finite-width model, let $p=xy_j+\delta_p$ be the rounded first product and let $\delta_y$ be the rounding error in the second product. The resulting residual is
\begin{equation}
r_{j+1}=r_j^2+xy_j\delta_p-x\delta_y.
\label{eq:nrround}
\end{equation}
Equation~(\ref{eq:nrround}) shows how a smaller exact seed residual can be hidden by intermediate rounding. Guard bits reduce the step size used for those products.

For Goldschmidt iteration, $n_0=y_0$ and $d_0=xy_0$. With the correction $c_j=2-d_j$, the updates are
\begin{equation}
n_{j+1}=n_jc_j,\qquad d_{j+1}=d_jc_j.
\label{eq:gs}
\end{equation}
In exact arithmetic the ratio $n_j/d_j=1/x$ is invariant in Equation~(\ref{eq:gs}). In fixed-point arithmetic, distinct rounding errors $\delta_n$ and $\delta_d$ are introduced. The ratio discrepancy $h_j=d_j-xn_j$ then satisfies
\begin{equation}
h_{j+1}=c_jh_j+\delta_d-x\delta_n.
\label{eq:gsround}
\end{equation}
The relative signed residual for the numerator is $1-xn_j=(1-d_j)+h_j$. Thus the update in Equation~(\ref{eq:gsround}) is needed in addition to the convergence of $d_j$. A broader finite-width analysis of Goldschmidt division is given by Even et al. \cite{EvenGoldschmidt05}. The final denominator product can be omitted when only the reciprocal is returned.

\subsection{Guard Bits and Residual Correction}\label{sec:residual}

The reconstruction uses the arithmetic widths recovered for each design. In the guarded alternatives an internal scale of $S_g=S2^g$ is used, where $g$ is the number of guard bits. The left endpoint code and the difference between rounded endpoint codes are stored. For a fractional position within the cell, the scaled difference is rounded at the internal scale. An optional Newton-Raphson step is evaluated as $y_0+Q_{F+g}(y_0[1-Q_{F+g}(xy_0)])$ before conversion to the output scale. This is the guarded form of Equation~(\ref{eq:nrround}) since $y_0$ is already on the internal grid.

For the residual-corrected alternatives, let $q_0$ be the candidate output code and $R=S^2-kq_0$ its signed integer residual. A one-step correction is
\begin{equation}
q_{\rm c}=
\begin{cases}
q_0+1,&2R\ge k,\\
q_0-1,&2R<-k,\\
q_0,&-k\le 2R< k.
\end{cases}
\label{eq:correct}
\end{equation}
If $|q_0-q^\star|\le 1$, the value from Equation~(\ref{eq:correct}) is $q^\star$. To verify this, the inequalities can be divided by $2k$ to locate the exact code $S^2/k$ relative to the half-code boundaries of the candidate. The final residual can also be checked using $-k\le 2(S^2-kq_{\rm c})<k$. The one-step condition was evaluated on both grids and the extra integer product was counted as part of the corrected datapath.

A sufficient condition can be derived for the guarded Hybrid. The rounded chord differs from the exact chord by at most $1/S_g$ because each endpoint and the interpolation product contribute at most half an internal LSB. Thus $|r_0|\le \Delta^2/[4a(a+\Delta)]+b/S_g$. The stored endpoint values and the interpolated value do not exceed 2. By Equation~(\ref{eq:nrround}), the ideal deviation of the rounded candidate in output LSB is bounded by
\begin{equation}
\begin{aligned}
\left|q_0-\frac{S^2}{k}\right|
&\le \frac{S}{a}\left(\frac{\Delta^2}{4a(a+\Delta)}+\frac{b}{S_g}\right)^2\\
&\quad+\frac{3}{2^{g+1}}+\frac12 .
\end{aligned}
\label{eq:candidate}
\end{equation}
If the right side of Equation~(\ref{eq:candidate}) is less than 1.5, then $|q_0-q^\star|\le 1$. The bound requires the stated nearest-rounding operations and enough product bits to avoid overflow.

\subsection{Ideal-Arithmetic Seed Planning}\label{sec:seed_planning}

Let $B_{\mathrm D}=\Delta/a$ and $B_{\mathrm C}=\Delta^2/[4a(a+\Delta)]$ denote continuous relative-error bounds for a left-endpoint table and a chord seed, respectively. The simpler bounds $B_{\mathrm D}=2^{2-N}$ and $B_{\mathrm C}\le 2^{2-2N}$ follow from the test interval (as shown in Section~\ref{sec:tables}). With a direct seed, $j$ steps of Newton-Raphson or Goldschmidt in exact arithmetic give a relative-error bound of $B_{\mathrm D}^{2^j}$. One Newton-Raphson step after a chord seed gives $B_{\mathrm C}^2$ (as shown in Section~\ref{sec:iteration}).

To compare seed widths, let $\tau_F(t)=t/S$ for $t\in\{a,b\}$ and let $L_F(t)=-\log_2\tau_F(t)$. If the ideal bound is at most $\tau_F(a)$, the ideal absolute error is at most one LSB at every input. The larger scale $\tau_F(b)$ expresses one LSB at the right endpoint only. The following sufficient address-width conditions are obtained by applying either scale to the simple bounds:
\begin{equation}
\begin{aligned}
N &\ge 2+L_F(t) &&\text{direct table},\\
N &\ge \tfrac12(2+L_F(t)) &&\text{chord alone},\\
N &\ge 2+L_F(t)/2^j &&\text{direct seed, $j$ refinements},\\
N &\ge \tfrac14(4+L_F(t)) &&\text{chord seed, one refinement}.
\end{aligned}
\label{eq:seed_planning}
\end{equation}

\begin{table}[H]
\noindent The address widths implied by Equation~(\ref{eq:seed_planning}) are compared in Table~\ref{tab:seed_planning}.

\smallskip
\caption{Address widths required by the stated continuous ideal-arithmetic bounds at each relative-error target.}
\label{tab:seed_planning}
\centering
\small
\begin{tabular}{@{}lrrrr@{}}
\toprule
& \multicolumn{2}{c}{Q16.16} & \multicolumn{2}{c}{Q8.24}\\
\cmidrule(lr){2-3}\cmidrule(l){4-5}
Method & $b/S$ & $a/S$ & $b/S$ & $a/S$\\
\midrule
Direct table & 17 & 19 & 25 & 27 \\
Interpolation & 9 & 10 & 13 & 14 \\
Direct seed and one refinement & 10 & 11 & 14 & 15 \\
Direct seed and two refinements & 6 & 7 & 8 & 9 \\
Interpolated seed and one refinement & 5 & 6 & 7 & 8 \\
\bottomrule
\end{tabular}
\end{table}

The sufficient uniform target is met in ideal arithmetic with six chord-seed address bits in Q16.16 and eight in Q8.24 when one refinement is used. These are the address widths evaluated for the guarded Hybrid (as shown in Section~\ref{sec:tuning}). The widths are obtained from continuous bounds. A direct-table width from this calculation may exceed the width needed to distinguish individual codes on the discrete input grid. Finite-width error, intermediate precision and correct rounding are evaluated separately.

\section{Evaluation Protocol}\label{sec:evaluation}

The reconstruction settings and external controls are defined for each arithmetic path (as shown in Sections~\ref{sec:reconstruction} and~\ref{sec:baselines}). The input coverage and reproducibility controls are then given (as shown in Section~\ref{sec:stimulus}). The RTL verification and synthesis procedure is described separately (as shown in Section~\ref{sec:hardwaremethod}).

\begin{figure}[H]
\noindent The verification paths are shown in Figure~\ref{fig:flow}.

\smallskip
  \centering
  \includegraphics[width=3.4in]{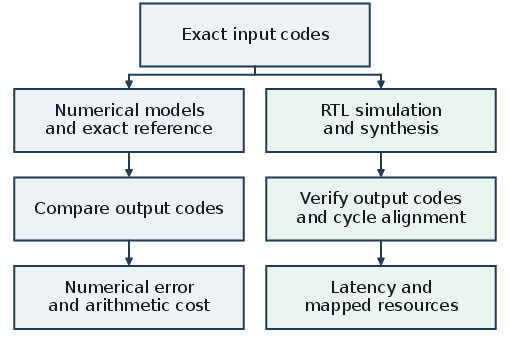}
  \Description{Exact input codes feed two verification paths. The numerical path compares output codes against an exact reference and records error and arithmetic cost. The hardware path checks output codes and cycle alignment and records latency and mapped resources.}
  \caption{Verification paths for the reciprocal arithmetic models and the hardware implementations.}
  \label{fig:flow}
\end{figure}

\subsection{Experiment Setup and Parameters}
\label{subsec:setup}

{Five fixed-point reciprocal architectures are evaluated: a direct lookup table (LUT), a lookup table with linear interpolation (INT), Newton-Raphson iteration (NR), Goldschmidt iteration (GS), and a hybrid (HYB) that seeds one Newton-Raphson step with an interpolated table value. Each is evaluated in two signed 32-bit formats, Q16.16 and Q8.24. Both formats have the same word length, so differences between them come from the position of the binary point. Guarded, residual-corrected and compact variants, the published baselines and an exact radix-2 control are evaluated on the same inputs.}

{The input interval is $[0.5, 2.5]$ with both endpoints. For $F$ fractional bits, every integer code from $2^{F-1}$ to $5\cdot 2^{F-1}$ is scored once, which gives 131{,}073 codes for Q16.16 and 33{,}554{,}433 codes for Q8.24. The grid is generated from integer codes and all arithmetic and references use exact integers, so the reported maxima are exact over the whole domain. A 1{,}000{,}001-point evenly spaced grid from the earlier experiment is also scored to show how much of the exhaustive maximum a sampled sweep captures.}

{Each output code $q$ for input code $u$ is compared with two references. The ideal error is $|q - 2^{2F}/u|$ in LSB, where one LSB is $2^{-F}$. The quantized reference is $2^{2F}/u$ rounded to the nearest code. We report the maximum ideal error, the maximum quantized error, the maximum relative error, the RMSE against each reference, the number of outputs equal to the quantized reference, and the smallest input code attaining each maximum. Zero maximum quantized error means the design is correctly rounded for every input. The numerical model also reports logical coefficient bits, multiplications, additions and table reads.}

{Four numerical studies are run on the exhaustive grids. The architecture comparison evaluates the five architectures at their reconstructed settings. The configuration sweep evaluates 153 configurations per format (306 in total), with 10 interpolation address widths from 5 to 14 and 7 hybrid address widths from 4 to 10, each with 9 guard widths from 0 to 8. The alternatives study evaluates smaller corrected hybrids and precision-matched Newton-Raphson and Goldschmidt variants. The baseline comparison evaluates the published quadratic seed, the degree-four and degree-ten polynomial corrections and the radix-2 control. The original coefficient tables were incomplete, so every ROM is regenerated from its defining equations, and the results describe this reconstruction.}

{Portable SystemVerilog is provided for the direct table, guarded interpolation, uncorrected and corrected Hybrid, and the unrolled radix-2 divider. Verilator streams every input code at one input per cycle and checks each output code and its cycle against the integer model. Yosys 0.52 maps each design to Xilinx 7 series primitives, and we report LUTs, flip-flops, DSP slices, block RAM in 36-kbit equivalents and CARRY4 cells. These are mapped resources and depend on the tool version. Routed timing, clock frequency, power and energy are not evaluated. The remaining subsections define each procedure, and Table~\ref{tab:setup} lists the parameters.}

\subsection{Reconstructed and Guarded Arithmetic}\label{sec:reconstruction}

The cited source revision \cite{KuttyArtifact} was used to identify address shifts, fixed-point products and pipeline organization. The finite-width implementation is described below (as shown in Section~\ref{sec:restored-datapaths}). The five reconstructed arithmetic settings use a 12-bit direct table, 11-bit chord interpolation, an 8-bit table and two Newton-Raphson steps, an 8-bit table and two Goldschmidt steps or a 9-bit interpolated seed and one Newton-Raphson step.

\begin{table}[H]
\noindent The reconstruction settings are summarized in Table~\ref{tab:designs}. The table includes logical operations on one input. Address formation, shifts and control are described separately from these counts.

\smallskip
\caption{Reconstructed arithmetic paths and the operations used for each input.}
\label{tab:designs}
\centering
\small
\begin{tabular}{@{}lrrrr@{}}
\toprule
Path & \shortstack[r]{Address\\bits} & \shortstack[r]{Refinement\\steps} & Products & \shortstack[r]{Table\\reads}\\
\midrule
Direct table & 12 & 0 & 0 & 1\\
Linear interpolation & 11 & 0 & 1 & 2\\
Newton-Raphson & 8 & 2 & 4 & 1\\
Goldschmidt & 8 & 2 & 4 & 1\\
Hybrid & 9 & 1 & 3 & 2\\
\bottomrule
\end{tabular}
\end{table}

Missing coefficient images were regenerated from the reciprocal at the tabulated breakpoints. For the recovered Q8.24 interpolation and the Hybrid, a signed slope with 21 fractional bits was formed from the ideal chord. For the recovered Q16.16 interpolation, the normalized cell phase and the product shift were used to infer a coefficient in output-code units. Its stored value was generated as the rounded ideal endpoint difference. The Q16.16 interpolation and Goldschmidt results depend on the regenerated coefficients and the selected rounding rules. These models reproduce the recovered arithmetic rules under the specified coefficient choices. The verification of the cycle behavior is limited to the separately identified portable RTL paths.

For the precision sweep, a shared endpoint interpolator and multiply rounding rule were used. The sweep included interpolation alone and interpolation with one Newton-Raphson step. The residual step was evaluated separately (as specified in Equation~(\ref{eq:correct})). The nominal read-only memory (ROM) payload is the number of entries multiplied by the width of the stored base and signed difference. The RTL can encode the difference as a nonnegative drop, saving one bit per entry. Mapped memory blocks depend on packing and widths.

\subsubsection{Functional Datapaths and Finite-Width Operations}\label{sec:restored-datapaths}

Let $H=2^{F+1-N}$ be the cell width in input codes. The address is formed from $k-S/2$ and is limited to the last address at the upper endpoint. For $k<5S/2$, the within-cell offset is $v=k-S/2-iH$. For the reconstructed interpolators at $k=5S/2$, the last address is used with an all-ones phase word. This differs from the guarded interpolation rule, which evaluates the right endpoint of the last chord (as described in Section~\ref{sec:residual}).

\begin{figure}[H]
\noindent The arithmetic dependencies of all five reconstructions are shown in Figure~\ref{fig:restored-architecture}. The direct path returns its stored base coefficient. The Hybrid passes its interpolated seed through one Newton-Raphson update. The Goldschmidt numerator and denominator updates in the first refinement are independent once their shared factor is formed.

\smallskip
\centering
\includegraphics[width=\linewidth]{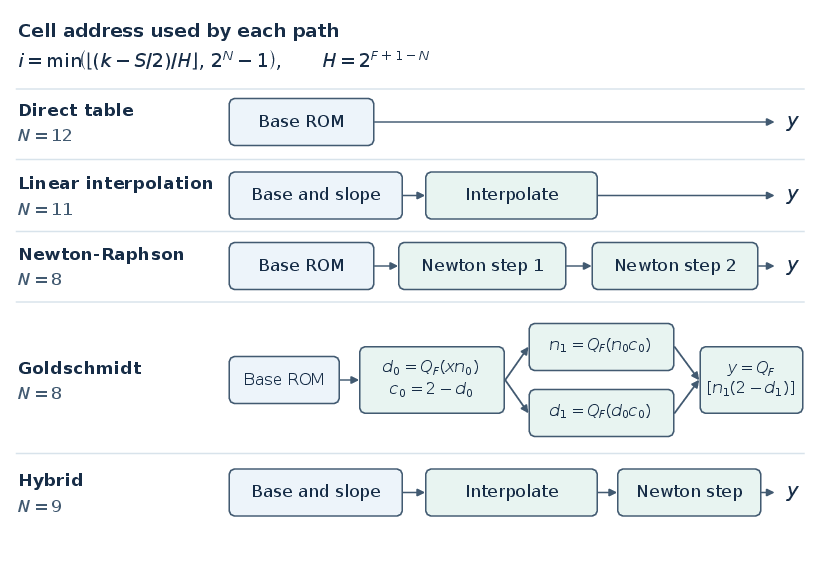}
\Description{Five rows show the reconstructed reciprocal paths after a common offset and clamp address rule. Direct lookup uses a 12-bit base table. Linear interpolation uses an 11-bit base and slope table. Newton-Raphson uses an 8-bit base table followed by two steps. Goldschmidt uses an 8-bit base table to initialize the denominator and has separate branches for the rounded numerator and denominator updates, followed by the final numerator. The Hybrid uses a 9-bit interpolated seed and one Newton-Raphson step.}
\caption{Functional dependencies of the five reconstructed reciprocal datapaths, with address widths from the evaluated numerical settings.}
\label{fig:restored-architecture}
\end{figure}

One Newton-Raphson step rounds $xy_j$ before forming $2-Q_F(xy_j)$ and rounds the second product to produce $y_{j+1}$ (as described in Equation~(\ref{eq:nrround})). In Goldschmidt refinement, the rounded products that form $n_1$ and $d_1$ can be evaluated in parallel (as described in Equation~(\ref{eq:gs})). The $d_1$ result is then used to form the correction for the final numerator $n_2$. Four multiplications are evaluated in total, with three sequential multiplication levels. A final denominator multiplication is unnecessary. These dependencies count arithmetic operations and do not specify the number of clock cycles.

\begin{table}[H]
\noindent The address shifts, stored fields and arithmetic used for the numerical comparison are given in Table~\ref{tab:restored-arithmetic}. The shift pairs are ordered as Q16.16 then Q8.24. The different coefficient encodings explain why the two linear interpolators cannot be described by a single signed shift.

\smallskip
\caption{Finite-width settings of the reconstructed arithmetic paths on the legal input interval.}
\label{tab:restored-arithmetic}
\centering
\small
\begin{tabular}{@{}>{\raggedright\arraybackslash}p{0.22\linewidth}>{\raggedright\arraybackslash}p{0.34\linewidth}>{\raggedright\arraybackslash}p{0.39\linewidth}@{}}
\toprule
Path and address shift & Stored fields and phase & Output arithmetic \\
\midrule
Direct table, 5 and 13
& One 32-bit base coefficient for each address
& Return the base coefficient without a product. \\
\addlinespace
Linear interpolation, 6 and 14
& 32-bit base and signed slope. Q16.16 uses an 18-bit endpoint difference in output-code units and an 11-bit phase. Q8.24 uses a 25-bit slope with 21 fractional bits and a 14-bit phase.
& Multiply the signed slope by the phase and shift the product arithmetically right by 11 or 21 bits, respectively. Add the shifted correction to the base. \\
\addlinespace
Newton-Raphson, 9 and 17
& One 32-bit base coefficient for each seed address
& Apply two refinements and round each product to $F$ fractional bits. \\
\addlinespace
Goldschmidt, 9 and 17
& One 32-bit base coefficient for each modeled seed address
& Round the four products individually to $F$ fractional bits. Return the last numerator. \\
\addlinespace
Hybrid, 8 and 16
& 32-bit base and signed 25-bit slope with 21 fractional bits. The phase is 8 or 16 bits, respectively.
& Shift the signed interpolation product right by 21 bits before adding it to the base. Apply one rounded Newton-Raphson step. \\
\bottomrule
\end{tabular}
\end{table}

For the reconstructed Q16.16 linear interpolator, the ordinary six-bit offset is shifted left by five bits to form the 11-bit phase. At the upper endpoint its phase is 2,047. The base is regenerated by rounding $S/x_i$ to an integer. For the Q16.16 linear interpolation case, the magnitude of the ideal endpoint difference in output codes is rounded and stored with a negative sign. For the other interpolators, the stored slope is the ideal chord slope scaled by $2^{21}$ with its magnitude rounded and its sign restored. The reconstructed outputs fit within the signed 32-bit range throughout the interval.

The public Q16.16 Goldschmidt module and its testbench use a nine-bit seed address with an eight-bit offset shift \cite{KuttyArtifact}. The numerical model instead uses the submitted manuscript's eight-bit seed address with a nine-bit shift to reproduce the specified comparison setting. This distinction and the missing original coefficient images limit any claim of equivalence to the final historical run. The independent portable RTL verification uses its own explicitly described design settings (as shown in Section~\ref{sec:hardwaremethod}).

\subsection{Published Methods and Exact Control}\label{sec:baselines}

The published Guti\'errez-Ramos et al. seed was evaluated using its six quadratic coefficient sets \cite{Gutierrez26}. For a given input, $x=2^e z$ with $z\in[1,2)$ and $e\in\{-1,0,1\}$. The coefficients of the selected segment are applied to $z$. The seed is refined using Equation~(\ref{eq:nr}) and the result is scaled by $2^{-e}$. A two-step variant using the same seed was included to show the effect of further refinement. The published coefficient codes have 17 fractional bits and were extended exactly to the common internal precision.

The polynomial correction of Lunglmayr was evaluated with the published degree-four and degree-ten coefficients \cite{Lunglmayr21}. For a polynomial $P_d$ of degree $d$, the output before final rounding was evaluated as
\begin{equation}
y=P_d(z-1)(3-z)2^{-1-e}.
\label{eq:lunglmayr}
\end{equation}
The polynomial in Equation~(\ref{eq:lunglmayr}) was evaluated by Horner's rule. Coefficients and internal products were rounded at $F+4$ fractional bits. The same internal product precision was applied to the quadratic baseline. The output was rounded after inverse scaling in both methods.

An exact radix-2 non-restoring control was evaluated on the same codes. Its integer quotient and remainder were used to form the rounded value of $S^2/k$ in Equation~(\ref{eq:codes}). The control implements the arithmetic recurrence. AMD PG151 describes an FPGA divider based on the same algorithm family \cite{AMD_PG151}. Published coefficients and algorithm rules were retained where applicable. The numerical ports share the input grid and output formats and use the stated arithmetic rules.

\subsection{Input Coverage and Scoring}\label{sec:stimulus}

The stimulus was generated from the exact integer interval $\mathcal K=\{S/2,\ldots,5S/2\}$, where both endpoints are included. Each code was scored once in the exhaustive runs. The evenly spaced, 1,000,001-point grid from the earlier experiment was also generated for the reconstructed paths. Its repeated Q16.16 codes and sampled Q8.24 codes were tracked independently.

\begin{table}[H]
\noindent The nominal parameters and the evidence used for evaluation are listed in Table~\ref{tab:setup}.

\smallskip
\caption{Parameters for the fixed-point reciprocal evaluation.}
\label{tab:setup}
\centering
\small
\begin{tabular}{@{}>{\raggedright\arraybackslash}p{0.27\linewidth}>{\raggedright\arraybackslash}p{0.69\linewidth}@{}}
\toprule
Parameter & Setting\\
\midrule
Input range & $[0.5,2.5]$ with both endpoints\\
Input and output & Signed 32-bit Q16.16 and Q8.24\\
Exhaustive test codes & 131,073 for Q16.16 and 33,554,433 for Q8.24\\
Earlier stimulus & 1,000,001 uniformly spaced real points rounded to the input grid\\
Reference & Integer reciprocal quotient with nearest rounding (Equation~\ref{eq:codes})\\
Accuracy metrics & Maximum relative error, maximum ideal absolute error, ideal RMSE and rounded-reference mismatches (Equations~\ref{eq:errors} and~\ref{eq:rmse})\\
Numerical cost & Logical coefficient bits, products, additions and table reads\\
Portable hardware flow & Verilator simulation and Yosys synthesis for the Xilinx 7 series cell library \\
\bottomrule
\end{tabular}
\end{table}

The dataset is synthetic and is specified fully by the code range. An existing online dataset was not used. The scripts regenerate the grid and include summary data, worst-case input codes and checks against the integer reference. These controls avoid dependence on a random permutation or on excluding a warm-up period for numerical scoring. For an RTL run, accepted transactions are tracked through reset and valid signals and pipeline fill and drain are distinguished from numerical coverage.

\subsection{Portable RTL Verification and Synthesis}\label{sec:hardwaremethod}

Portable synchronous RTL was generated for direct lookup, guarded interpolation, guarded Hybrid refinement, residual correction and exact radix-2 division. The direct and interpolation address widths were 12 and 11 bits, respectively. The default refined designs used four guard bits and an address width of six bits for Q16.16 or eight bits for Q8.24. The exact radix-2 datapath uses $F+2$ unrolled recurrence steps. Separate register stages were used for its final remainder and rounding operations.

The full-product corrector computes $kq_0$ directly. In the compact corrector, let $Z$ be the original guarded seed code and let $D=2^gq_0-Z$. The full product $kZ$ is already available from refinement. The scaled residual is recovered as
\begin{equation}
2^gR=(S^2\,2^g-kZ)-kD .
\label{eq:reuse}
\end{equation}
The remaining product in Equation~(\ref{eq:reuse}) has the narrower signed operand $D$. The thresholds in Equation~(\ref{eq:correct}) are scaled by $2^g$. In the compact datapath, the final input endpoint is handled by subtracting the full tabulated drop. This permits a narrower cell-phase operand. Four guard bits were used for a matched comparison of the full-product and compact designs. The compact design was also evaluated at $g=0$ for Q16.16 and at $g=1$ and $g=2$ for Q8.24. Their one-step correction conditions were checked exhaustively.

A C++ scoreboard was run with Verilator 5.48.0 \cite{VerilatorDocs}. The output code and its cycle were checked against an independent arithmetic model for every accepted input. The entire domain was streamed with an initiation interval of one cycle. Directed tests checked valid gaps and reset with transactions in flight. Each design was mapped with the same synthesis command in Yosys 0.52 using the \texttt{xc7} technology library \cite{YosysXilinx}. A lookup table (LUT) count refers to the mapped FPGA logic cells, while logical coefficient bits refer to the stored numerical values. Counts are given for flip-flops (FFs), digital signal processing (DSP) slices and block random-access memory (BRAM), with BRAM reported in 36-kilobit equivalents. The 7 series DSP slice ports and arithmetic widths are described in the device guide \cite{XilinxUG479}.

\section{Results}\label{sec:results}

The effects of format and architecture choices address the two research questions (as stated in Section~\ref{sec:intro}). The recovered arithmetic and the coverage of the earlier sweep are evaluated first (as shown in Section~\ref{sec:reconstructionresults}). Error distributions and the cross-format error budget are examined next (as shown in Sections~\ref{sec:errordistribution} and~\ref{sec:error_format}). Changes in seed width and guard precision and the resulting choices under explicit accuracy budgets are then compared (as shown in Sections~\ref{sec:tuning} and~\ref{sec:numericalbudget}). Published polynomial methods and the exact control are evaluated under the same scoring rules (as shown in Section~\ref{sec:comparison}). The portable RTL results and architecture selections under hardware constraints are then reported (as shown in Sections~\ref{sec:hardwareresults} and~\ref{sec:portable_tradeoffs}).

\subsection{Reconstructed Designs and Input Coverage}\label{sec:reconstructionresults}

For the reconstructed Q8.24 Hybrid, the maximum ideal deviation was 1.4979 LSB over the complete grid and the maximum rounded-code deviation was one LSB. The exhaustive direct-table maximum of 32,732.04 LSB exceeded the 32,716.07 LSB observed in the earlier style of sweep. Each Q16.16 code was visited by the million-point stimulus. The Q8.24 stimulus covered 2.98\% of the available codes.

\begin{table}[H]
\noindent Table~\ref{tab:reconstruction} reports the maximum and average errors against the ideal and rounded references on the exhaustive grids. The rounded-code RMSE and mean relative error are computed as shown in Equation~(\ref{eq:aggregate_error}).

\smallskip
\caption{Exhaustive numerical accuracy of the five reconstructed arithmetic paths.}
\label{tab:reconstruction}
\centering
\small
\setlength{\tabcolsep}{2.6pt}
\begin{tabular}{@{}llrrrrrr@{}}
\toprule
& & \multicolumn{4}{c}{Ideal reference} & \multicolumn{2}{c}{Rounded reference}\\
\cmidrule(lr){3-6}\cmidrule(l){7-8}
Format & Path & \shortstack[r]{Max. rel.\\(\%)} & \shortstack[r]{Mean rel.\\(\%)} & \shortstack[r]{Max. abs.\\(LSB)} & \shortstack[r]{RMSE\\(LSB)} & \shortstack[r]{Max. abs.\\(LSB)} & \shortstack[r]{RMSE\\(LSB)}\\
\midrule
Q16.16 & Direct table & 0.094604 & 0.019055 & 123.883 & 20.754 & 124 & 20.755 \\
 & Interpolation & 0.006517 & 0.001235 & 1.961 & 0.647 & 2 & 0.708 \\
 & Newton-Raphson & 0.002289 & \(6.7523\times 10^{-4}\) & 1.462 & 0.381 & 1 & 0.451 \\
 & Goldschmidt & 0.004123 & \(8.3709\times 10^{-4}\) & 2.535 & 0.507 & 3 & 0.569 \\
 & Hybrid & 0.002289 & \(6.5413\times 10^{-4}\) & 1.245 & 0.357 & 1 & 0.418 \\
\addlinespace[2pt]
Q8.24 & Direct table & 0.097644 & 0.019649 & 32,732.039 & 5,439.956 & 32,732 & 5,439.956 \\
 & Interpolation & \(9.5529\times 10^{-5}\) & \(1.1778\times 10^{-5}\) & 31.961 & 5.022 & 32 & 5.030 \\
 & Newton-Raphson & \(1.0391\times 10^{-5}\) & \(2.6537\times 10^{-6}\) & 3.410 & 0.403 & 3 & 0.463 \\
 & Goldschmidt & \(2.0117\times 10^{-5}\) & \(3.6136\times 10^{-6}\) & 4.221 & 0.554 & 4 & 0.624 \\
 & Hybrid & \(1.0248\times 10^{-5}\) & \(2.6318\times 10^{-6}\) & 1.498 & 0.388 & 1 & 0.450 \\
\bottomrule
\end{tabular}
\end{table}

\begin{figure}[H]
\noindent The extent to which the 1,000,001-point sweep captured the maxima is shown in Figure~\ref{fig:coverage}. Each pair of bars was computed using the same arithmetic model.

\smallskip
  \centering
  \includegraphics[width=3.4in]{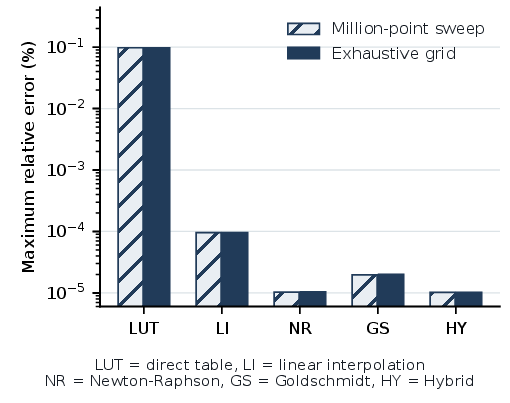}
  \Description{Paired bars on a logarithmic scale compare maximum relative error from the million-point sweep and exhaustive Q8.24 grid for the direct lookup, linear interpolation, Newton-Raphson, Goldschmidt and Hybrid paths.}
  \caption{Observed and exhaustive maximum relative errors for the reconstructed Q8.24 paths.}
  \label{fig:coverage}
\end{figure}

The released accuracy summary and the later historical experiment refer to different runs. The final coefficients and raw outputs were unavailable at the cited revision. The results in Table~\ref{tab:reconstruction} were obtained with the stated reconstruction and its regenerated coefficient images. The input codes at the maxima are included in the result files.

\subsection{Error Distributions and Input Dependence}\label{sec:errordistribution}

The signed difference $q-q^\star$ was counted at every input code for each reconstructed architecture. Absolute-error percentiles were then obtained by selecting the smallest magnitude whose cumulative count reached the stated percentage. Each input contributed once. The signed distribution describes agreement with the rounded reference. The input profiles use the errors against the ideal reciprocal defined in Equation~(\ref{eq:errors}).

\begin{figure}[H]
\noindent The signed rounded-code errors and their absolute-error percentiles are shown in Figure~\ref{fig:reconstructed_distribution}.

\smallskip
  \centering
  \includegraphics[width=\linewidth]{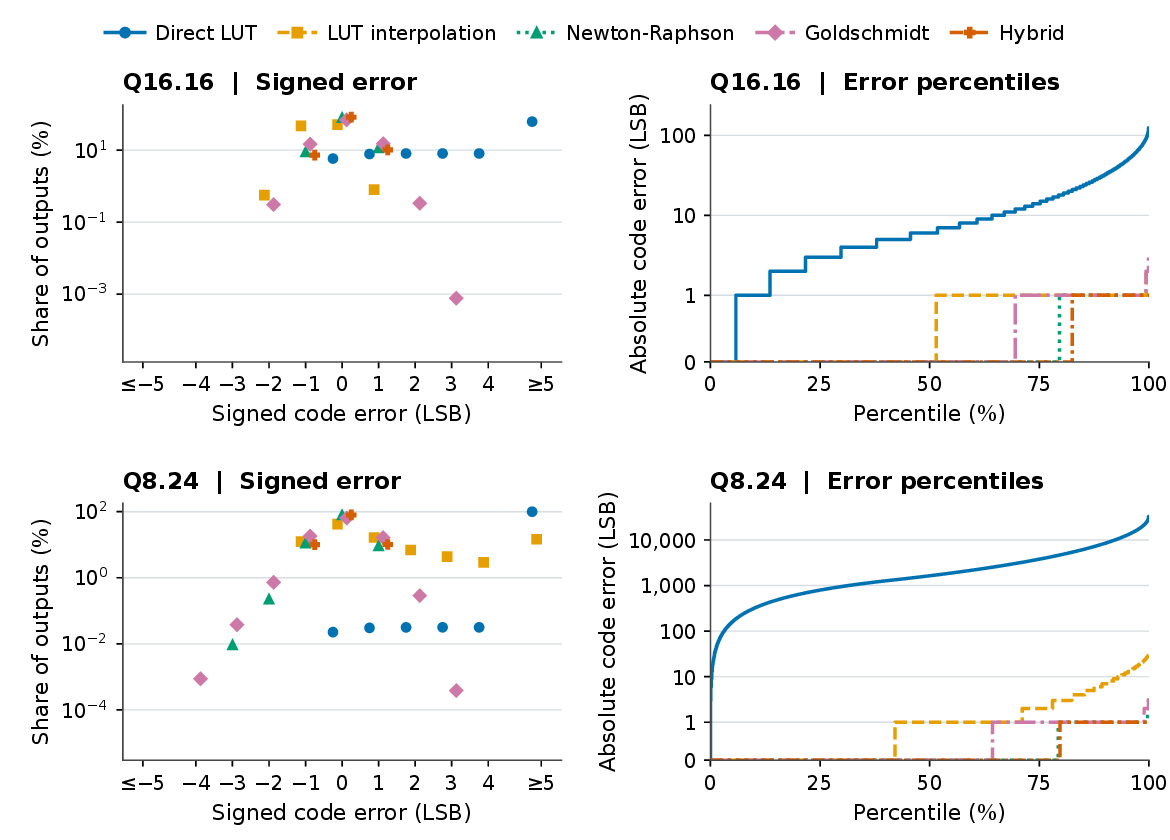}
  \Description{For each fixed-point format the left panel shows the percentage of outputs in each signed code error category and the right panel shows absolute code error percentiles. The direct table has only nonnegative rounded-code errors and the Hybrid stays within one code of the rounded reference in both formats.}
  \caption{Rounded-reference error distributions for the five reconstructed paths on the exhaustive Q16.16 and Q8.24 grids, with signed errors outside $[-4,4]$ grouped into two tails and zero-count categories omitted.}
  \label{fig:reconstructed_distribution}
\end{figure}

The direct lookup has no negative rounded-code differences because the rounded table value at the left endpoint is at least the rounded reciprocal at any input in that cell. For the reconstructed Q16.16 interpolator, 47.69\% of the outputs are below the rounded reference and 0.80\% are above it. This asymmetry is specific to the reconstructed coefficients and arithmetic shifts. The Hybrid stays within one output code in both formats. In Q8.24 the 99th percentile is one LSB for Newton-Raphson and the Hybrid and two LSB for Goldschmidt. At the 95th percentile the Q8.24 direct lookup differs by 12{,}703 LSB and interpolation differs by 12 LSB.

\begin{figure}[H]
\noindent The variation of the maximum ideal errors with the input is shown in Figure~\ref{fig:reconstructed_profiles}.

\smallskip
  \centering
  \includegraphics[width=\linewidth]{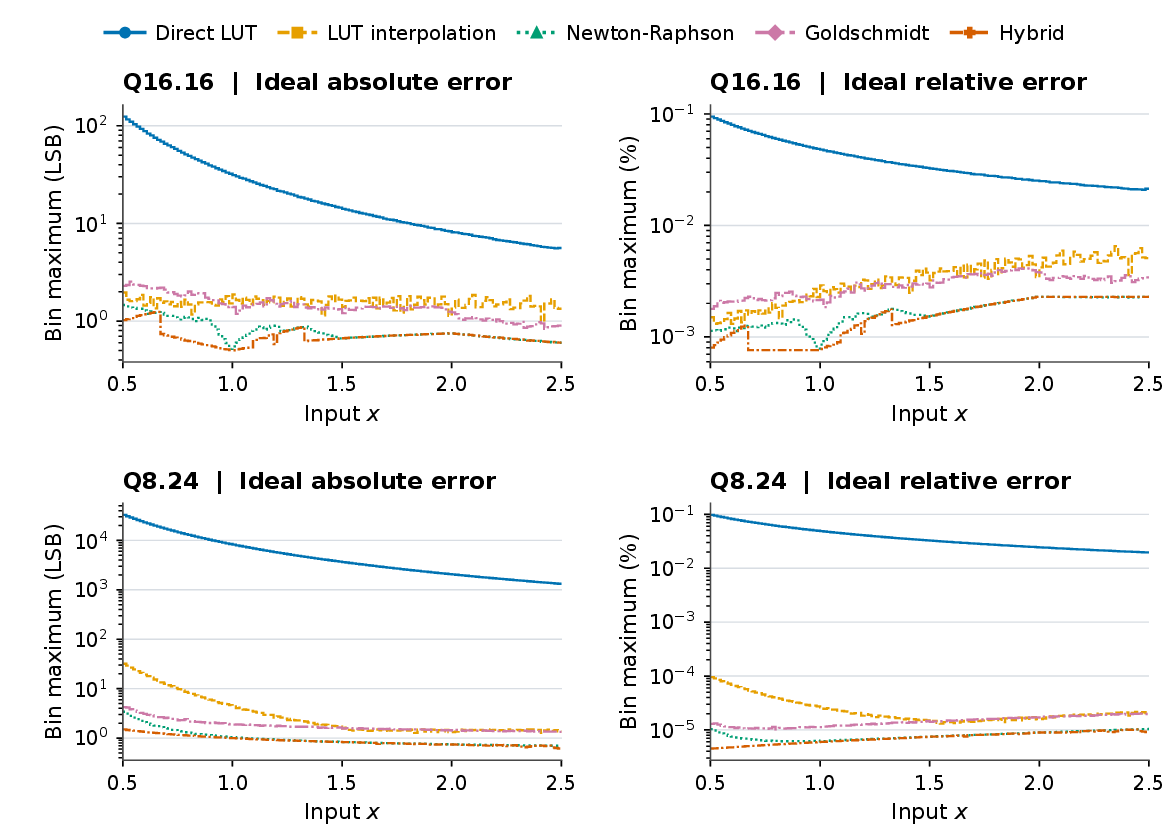}
  \Description{Two panels per format show maximum ideal absolute and relative errors in 128 input bins. Both maxima of the Q8.24 direct lookup and interpolation occur in the first bin. The Q8.24 Hybrid has its largest absolute error in the first bin and its largest relative error in a bin near the upper end of the interval.}
  \caption{Maximum ideal absolute and relative errors within 128 equal-width bins for the five reconstructed paths, evaluated over every input code with the right endpoint included in the final bin.}
  \label{fig:reconstructed_profiles}
\end{figure}

The first Q8.24 bin contains both global maxima for the direct lookup and interpolation. It also contains the largest ideal absolute error for the Hybrid, while its largest ideal relative error occurs in the bin beginning at $x=2.4375$. For each input, Equation~(\ref{eq:errors}) gives $E_{\rm rel}=100xE_{\rm abs}/S$. The factor $x$ explains why the two measures can reach their maxima in different bins. The signed counts and bin maxima are supplied with the reproduction files. Transformed histograms and conditional mismatch quartiles are given in Appendix~\ref{app:error_distributions}.

\subsection{Error Budgets and Format Sensitivity}\label{sec:error_format}

For the direct table, let $d=\Delta-1/S$ be the last interior displacement on the input grid. Both errors in Equation~(\ref{eq:direct}) increase with displacement and decrease with the left cell boundary. The largest interior error is therefore attained in the first cell. The right endpoint is considered separately because its displacement is $\Delta$. With exact table values, the resulting grid maxima are
\begin{equation}
\begin{aligned}
E_{\rm rel}^{\rm grid} &=100\max\left\{\frac{d}{a},\frac{\Delta}{b-\Delta}\right\},\\
E_{\rm abs}^{\rm grid} &= S\max\left\{\frac{d}{a(a+d)},\frac{\Delta}{b(b-\Delta)}\right\}.
\end{aligned}
\label{eq:direct_grid}
\end{equation}
The reconstructed direct table stores the value $2$ exactly in the first cell, where both observed maxima occur. Its final cell is retained in the check even though neither maximum is attained there.

\begin{table}[H]
\noindent Equation~(\ref{eq:direct_grid}) is checked against the complete input grids in Table~\ref{tab:direct_grid}.

\smallskip
\caption{Direct-table grid predictions and exhaustive ideal-reference maxima for the 12-bit address.}
\label{tab:direct_grid}
\centering
\small
\begin{tabular}{@{}lrrrr@{}}
\toprule
& \multicolumn{2}{c}{Relative error (\%)} & \multicolumn{2}{c}{Absolute error (LSB)}\\
\cmidrule(lr){2-3}\cmidrule(l){4-5}
Format & Grid prediction & Observed & Grid prediction & Observed\\
\midrule
Q16.16 & 0.094604492 & 0.094604492 & 123.882801 & 123.882801 \\
Q8.24 & 0.097644329 & 0.097644329 & 32732.039020 & 32732.039020 \\
\bottomrule
\end{tabular}
\end{table}

\begin{table}[H]
\noindent The ideal approximation bounds and exhaustive relative errors are compared in Table~\ref{tab:ideal_budget}. The ideal bounds assume exact coefficients and products.

\smallskip
\caption{Continuous ideal-arithmetic bounds and observed maximum relative errors for the reconstructed paths.}
\label{tab:ideal_budget}
\centering
\small
\begin{tabular}{@{}lrrrr@{}}
\toprule
Path & $N$ & Ideal bound (\%) & Q16.16 (\%) & Q8.24 (\%)\\
\midrule
Direct table & 12 & 0.097656 & 0.094604 & 0.097644 \\
Interpolation & 11 & \(9.5182\times 10^{-5}\) & 0.006517 & \(9.5529\times 10^{-5}\) \\
Newton-Raphson & 8 & \(5.9605\times 10^{-6}\) & 0.002289 & \(1.0391\times 10^{-5}\) \\
Goldschmidt & 8 & \(5.9605\times 10^{-6}\) & 0.004123 & \(2.0117\times 10^{-5}\) \\
Hybrid & 9 & \(2.2923\times 10^{-8}\) & 0.002289 & \(1.0248\times 10^{-5}\) \\
\bottomrule
\end{tabular}
\end{table}

The direct-table errors agree with the discrete prediction to the displayed precision. The Q8.24 interpolation maximum is close to the ideal chord bound. The Q16.16 interpolation maximum and the maxima of the iterative and Hybrid paths exceed their ideal bounds. For the interpolated and refined paths the ideal bounds lie below the quantization floor of the output format. A correctly rounded output can have a relative error of up to $50b/S$ percent, which is $0.0019\%$ in Q16.16 and $7.4\times10^{-6}\%$ in Q8.24, so the observed maxima in Table~\ref{tab:ideal_budget} should be compared with this floor and not with the ideal bound. Exceeding the bound there reflects output rounding and not a deficiency of the approximation. Differences between the two columns do not isolate the magnitude of coefficient, product and output errors because these errors can reinforce or offset one another.

\begin{table}[H]
\noindent The effect of fractional precision is summarized in Table~\ref{tab:cross_format}. The first ratio is the Q16.16 maximum relative error divided by the Q8.24 maximum relative error. The scale ratios divide each maximum relative error by the percentage equivalent of $b/S$ in that format.

\smallskip
\caption{Exhaustive error ratios and agreement with the rounded reference for the two formats.}
\label{tab:cross_format}
\centering
\small
\setlength{\tabcolsep}{3.2pt}
\begin{tabular}{@{}lrrrrr@{}}
\toprule
& \shortstack{Max. rel.\\ratio} & \multicolumn{2}{c}{Scale ratio} & \multicolumn{2}{c}{Exact share (\%)}\\
\cmidrule(lr){3-4}\cmidrule(l){5-6}
Path & Q16.16/Q8.24 & Q16.16 & Q8.24 & Q16.16 & Q8.24\\
\midrule
Direct table & 0.969 & 24.8 & 6553 & 5.846 & 0.023 \\
Interpolation & 68.2 & 1.708 & 6.411 & 51.511 & 42.171 \\
Newton-Raphson & 220 & 0.6 & 0.6974 & 79.643 & 79.354 \\
Goldschmidt & 205 & 1.081 & 1.35 & 69.511 & 64.386 \\
Hybrid & 223 & 0.6 & 0.6877 & 82.521 & 79.737 \\
\bottomrule
\end{tabular}
\end{table}

\begin{figure}[H]
\noindent The worst deviations from the ideal and rounded references are compared in Figure~\ref{fig:error_budget}.

\smallskip
\centering
\includegraphics[width=\linewidth]{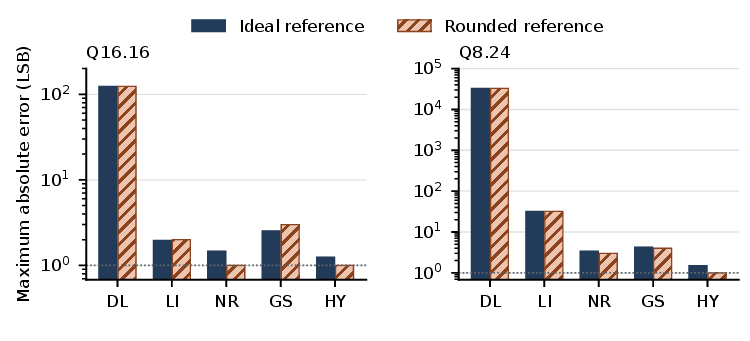}
\Description{Grouped logarithmic bars compare maximum ideal absolute error and maximum rounded-reference error for five reconstructed designs in Q16.16 and Q8.24. A line marks one output LSB.}
\caption{Exhaustive maximum error against each reference in units of the format-specific output LSB, with the dotted line at one LSB. DL denotes direct lookup, LI linear interpolation, NR Newton-Raphson, GS Goldschmidt and HY Hybrid.}
\label{fig:error_budget}
\end{figure}

The exact share is the fraction of tested codes with $E_{\rm code}=0$, expressed as a percentage. An error below the $b/S$ scale does not imply that every code is within one LSB of the ideal reciprocal. The one-LSB relative threshold depends on the input, as in Equation~(\ref{eq:seed_planning}).

The eight extra fractional bits lower the worst relative error by about $68$ times for the reconstructed interpolator and by $205$ to $223$ times for the three refined paths. The direct-table maximum remains close to $0.1\%$ because the address width is unchanged. The Q8.24 Hybrid also has the smallest ideal absolute maximum among the reconstructed Q8.24 paths. Its exact share of $79.737\%$ quantifies agreement at the finer output grid. Exact share and maximum absolute error describe different aspects of an approximation.

\begin{figure}[H]
\noindent Relative-error maxima and root mean square errors in common reciprocal units are shown in Figure~\ref{fig:cross_format}.

\smallskip
\centering
\includegraphics[width=\linewidth]{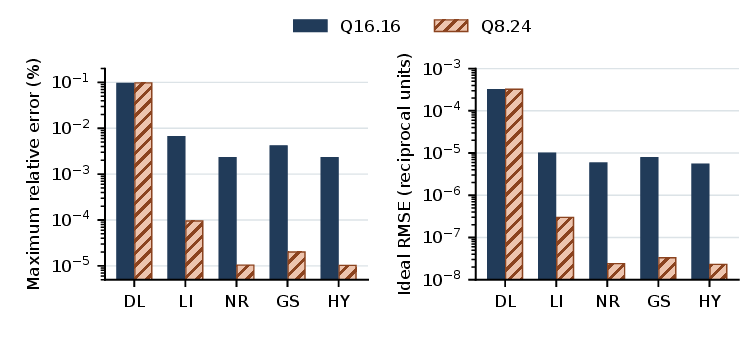}
\Description{The two formats are compared in grouped logarithmic bars for maximum ideal relative error and ideal RMSE rescaled to reciprocal units. Direct-table error is nearly unchanged. Interpolation, Newton-Raphson, Goldschmidt and Hybrid have lower absolute error in Q8.24.}
\caption{Cross-format comparison on complete grids, with ideal RMSE divided by $S$ to express it in common reciprocal units. The architecture abbreviations are defined in Figure~\ref{fig:error_budget}.}
\label{fig:cross_format}
\end{figure}

\subsection{Seed Width and Guard Precision}\label{sec:tuning}

The sweep included 153 combinations of interpolation width, guard precision and refinement per format (306 in total). With four guard bits, the selected Hybrid candidates reached 0.5973 LSB in Q16.16 and 0.6462 LSB in Q8.24 before correction. The sufficient bounds in Equation~(\ref{eq:candidate}) are approximately 0.705 and 0.715 LSB, respectively. The subsequent residual correction yielded zero rounded-code mismatches in both formats. The smallest tested coefficient payloads with a precursor within one rounded-code LSB were 2,048 and 12,288 bits for Q16.16 and Q8.24, respectively. These compact precursors are also sufficient for exact correction. The maximum ideal error and the number of rounded-reference mismatches show distinct effects of internal precision.

\begin{table}[H]
\noindent Selected operating points are given in Table~\ref{tab:tuning}.

\smallskip
\caption{Selected guarded interpolation and refinement configurations on the exhaustive input grids.}
\label{tab:tuning}
\centering
\small
\begin{tabular}{@{}llrrrrr@{}}
\toprule
Format & Method & $N$ & $g$ & \shortstack[r]{ROM\\bits} & \shortstack[r]{Ideal max.\\(LSB)} & Mismatches\\
\midrule
Q16.16 & Guarded interpolation & 11 & 4 & 71,680 & 0.6640 & 5,625 \\
 & Guarded hybrid & 6 & 4 & 2,560 & 0.5973 & 4,048 \\
 & Corrected hybrid & 6 & 4 & 2,560 & 0.5000 & 0 \\
 & Compact precursor & 6 & 0 & 2,048 & 1.4485 & 27,442 \\
\addlinespace[2pt]
Q8.24 & Guarded interpolation & 11 & 4 & 104,448 & 32.4356 & 22,258,973 \\
 & Guarded hybrid & 8 & 4 & 13,824 & 0.6462 & 1,027,084 \\
 & Corrected hybrid & 8 & 4 & 13,824 & 0.5000 & 0 \\
 & Compact precursor & 8 & 1 & 12,288 & 1.2379 & 8,465,471 \\
\bottomrule
\end{tabular}
\end{table}

\begin{figure}[H]
\noindent The relation between logical memory and maximum ideal error is shown in Figure~\ref{fig:tuning}. An ideal deviation of at most half an LSB follows from correct rounding. The exact match counts are used to check this property on the input grid.

\smallskip
  \centering
  \includegraphics[width=3.4in]{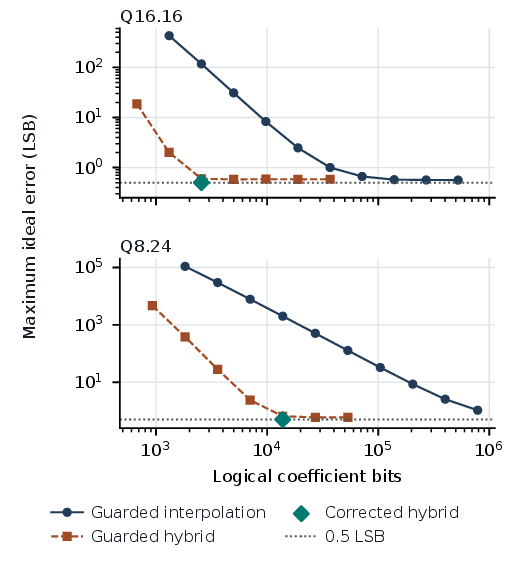}
  \Description{Two logarithmic plots for Q16.16 and Q8.24 show maximum ideal error against logical coefficient bits for guarded interpolation and guarded Hybrid refinement. A marker shows each corrected Hybrid and a horizontal line marks half an output LSB.}
  \caption{Effect of table size on guarded interpolation and multiplicative refinement.}
  \label{fig:tuning}
\end{figure}

Corrected interpolation uses the residual check after a wider table with no refinement. The two-code correction also checks $2R$ against $3k$ and $-3k$. It adds two codes if $2R\ge 3k$ and subtracts two if $2R<-3k$. Otherwise the rule in Equation~(\ref{eq:correct}) is used. This rule gives the correctly rounded result when the candidate differs from the reference by at most two codes. The two-code path uses smaller tables and requires two extra thresholds. Synthesis and cycle measurements are supplied for the compact one-code alternative (as shown in Section~\ref{sec:hardwareresults}).

\begin{table}[H]
\noindent Table~\ref{tab:alternatives} lists three further configurations that returned the rounded reference for every input. Together with the corrected Hybrid of Table~\ref{tab:tuning}, this gives four exactly rounded table configurations per format.

\smallskip
\caption{Numerically exact alternatives and their signed-difference coefficient payloads.}
\label{tab:alternatives}
\centering
\small
\begin{tabular}{@{}llrrrr@{}}
\toprule
Format & Method & $N$ & $g$ & ROM bits & Products\\
\midrule
Q16.16 & Compact corrected hybrid & 6 & 0 & 2,048 & 4 \\
 & Corrected interpolation & 10 & 0 & 28,672 & 2 \\
 & Two-code corrected hybrid & 5 & 1 & 1,120 & 4 \\
\addlinespace[2pt]
Q8.24 & Compact corrected hybrid & 8 & 1 & 12,288 & 4 \\
 & Corrected interpolation & 14 & 0 & 655,360 & 2 \\
 & Two-code corrected hybrid & 7 & 2 & 6,528 & 4 \\
\bottomrule
\end{tabular}
\end{table}

Two-step Newton-Raphson and Goldschmidt alternatives were also evaluated with the same guard precision as the guarded Hybrid. At the Hybrid address width their direct seeds led to maximum rounded-reference differences of two codes. With one extra address bit both remained within one code on both input grids. Their complete results are included with the width sweep.

\subsection{Numerical Selection Under Accuracy Constraints}\label{sec:numericalbudget}

The reconstructed, guarded and residual-corrected table designs were compared under common accuracy requirements. The maximum ideal error $E_{\max}=\max_{k\in\mathcal K}E_{\rm abs}(k)$ was used for the bounded-error requirements. Exact rounding required $E_{\rm code}(k)=0$ for every input code. The selection covered 166 distinct configurations in each format, namely the 153 swept configurations together with the reconstructed, corrected and compact designs, with repeated CSV entries for the same configuration counted once. For each requirement the least coefficient memory and fewest products were found independently. Ties in the primary objective were resolved using the other cost and then $E_{\max}$. The digit-recurrence control is characterized by its recurrence steps (Table~\ref{tab:baseline}).

\begin{table}[H]
\noindent Table~\ref{tab:numericalbudget} gives the two selections for each requirement. Each configuration is identified by its seed width and guard precision $(N,g)$. LI denotes reconstructed linear interpolation, GI guarded interpolation, GH guarded Hybrid, CI corrected interpolation and H2 the two-code corrected Hybrid.

\smallskip
\caption{Minimum logical coefficient memory and minimum arithmetic product count among the evaluated table designs meeting each accuracy requirement.}
\label{tab:numericalbudget}
\centering
\small
\begin{tabular}{@{}ll lrr lrr@{}}
\toprule
& & \multicolumn{3}{c}{Least coefficient memory} & \multicolumn{3}{c}{Fewest products} \\
\cmidrule(lr){3-5}\cmidrule(l){6-8}
Format & Requirement & Design & Bits & Products & Design & Bits & Products \\
\midrule
Q16.16 & $\leq 32$ LSB & GH $(4,0)$ & 544 & 3 & GI $(7,0)$ & 3,968 & 1 \\
 & $\leq 4$ LSB & GH $(5,0)$ & 1,056 & 3 & GI $(9,0)$ & 14,848 & 1 \\
 & $\leq 1$ LSB & H2 $(5,1)$ & 1,120 & 4 & GI $(10,5)$ & 38,912 & 1 \\
 & Exact rounding & H2 $(5,1)$ & 1,120 & 4 & CI $(10,0)$ & 28,672 & 2 \\
\addlinespace[2pt]
Q8.24 & $\leq 32$ LSB & GH $(6,0)$ & 3,072 & 3 & LI $(11,0)$ & 116,736 & 1 \\
 & $\leq 4$ LSB & GH $(7,0)$ & 6,016 & 3 & GI $(13,0)$ & 335,872 & 1 \\
 & $\leq 1$ LSB & H2 $(7,2)$ & 6,528 & 4 & GI $(14,8)$ & 917,504 & 1 \\
 & Exact rounding & H2 $(7,2)$ & 6,528 & 4 & CI $(14,0)$ & 655,360 & 2 \\
\bottomrule
\end{tabular}
\end{table}

With a four-LSB budget, the three-product guarded Hybrid used 1,056 coefficient bits in Q16.16 and 6,016 bits in Q8.24. The corresponding one-product alternatives needed 14,848 and 335,872 bits. A one-LSB budget could still be met with one product in both formats, although the Q8.24 interpolator required 917,504 bits. For exact rounding, the two-product corrected interpolator used 28,672 bits in Q16.16 and 655,360 bits in Q8.24. The four-product two-code Hybrid returned the same output codes with 1,120 and 6,528 bits, respectively. It includes the extra threshold comparisons described in Section~\ref{sec:tuning}.

Strict dominance was also evaluated over maximum ideal error, coefficient bits and products per result. A configuration dominated another if it was at least as good on all three metrics and strictly better on at least one. Among the four exactly rounded table configurations tested per format, the two-product corrected interpolator and four-product two-code Hybrid were the two non-dominated choices in both formats. The separately mapped implementations in Section~\ref{sec:hardwareresults} provide cycle and resource measurements for the hardware candidates.

Sensitivity to the reporting precision was assessed by replacing $E_{\max}$ in the dominance comparison with the integer class $\lceil E_{\max}/\delta\rceil$, for class width $\delta$ in LSB. Errors assigned to the same class were treated as equal. This rule gives a transitive equivalence relation, with boundaries at integer multiples of $\delta$. Coefficient memory and product count were compared at full precision throughout. Exact rounding was assessed from the mismatch count.

\begin{table}[H]
\noindent Table~\ref{tab:numericalparetosensitivity} shows how the non-dominated set changes with the resolution of the ideal-error objective.

\smallskip
\caption{Number of non-dominated table configurations under alternative ideal-error resolutions.}
\label{tab:numericalparetosensitivity}
\centering
\small
\begin{tabular}{@{}lrr@{}}
\toprule
Ideal-error comparison & Q16.16 & Q8.24 \\
\midrule
Unrounded maximum & 56 & 91 \\
Classes of 0.25 LSB & 24 & 36 \\
Classes of 0.5 LSB & 19 & 33 \\
Classes of 1 LSB & 15 & 24 \\
\bottomrule
\end{tabular}
\end{table}

The coarser classes allowed a lower-cost configuration to dominate an alternative whose error improvement fell within the same class. Within the five reconstructed designs, Newton-Raphson dominated Goldschmidt under the unrounded objectives in both formats. Both used 8,192 coefficient bits and four products, while Newton-Raphson had the lower maximum ideal error in both grids (Table~\ref{tab:reconstruction}). This was the sole strict dominance relationship among the reconstructed designs under those objectives.

\subsection{Comparison With Published Methods}\label{sec:comparison}

\begin{table}[H]
\noindent The numerical comparison under the common format and scoring rules is shown in Table~\ref{tab:baseline}. In the table and figure labels, NR denotes Newton-Raphson. For the polynomial methods the multiplication count excludes exact power-of-two scaling and addition. The digit-recurrence control is characterized by recurrence steps. A product count does not predict the latency or DSP count of a placed design.

\smallskip
\caption{Exhaustive accuracy and arithmetic effort of the published-method baselines and selected guarded alternatives.}
\label{tab:baseline}
\centering
\small
\begin{tabular}{@{}llrrrr@{}}
\toprule
Format & Method & \shortstack[r]{Ideal max.\\(LSB)} & \shortstack[r]{RMSE\\(LSB)} & Mismatches & Products\\
\midrule
Q16.16 & Quadratic + 1 NR & 0.6209 & 0.2915 & 4,452 & 5 \\
 & Quadratic + 2 NR & 0.5936 & 0.2917 & 4,734 & 7 \\
 & Lunglmayr d4 & 6.7074 & 2.1430 & 111,795 & 5 \\
 & Lunglmayr d10 & 0.6166 & 0.2896 & 2,303 & 11 \\
 & Exact radix-2 & 0.5000 & 0.2886 & 0 & 0 \\
 & Guarded hybrid & 0.5973 & 0.2906 & 4,048 & 3 \\
 & Corrected hybrid & 0.5000 & 0.2886 & 0 & 4 \\
\addlinespace[2pt]
Q8.24 & Quadratic + 1 NR & 4.4184 & 0.7954 & 9,341,442 & 5 \\
 & Quadratic + 2 NR & 0.6133 & 0.2921 & 1,263,590 & 7 \\
 & Lunglmayr d4 & 1663.0002 & 543.6319 & 33,535,744 & 5 \\
 & Lunglmayr d10 & 0.7106 & 0.2906 & 827,959 & 11 \\
 & Exact radix-2 & 0.5000 & 0.2887 & 0 & 0 \\
 & Guarded hybrid & 0.6462 & 0.2906 & 1,027,084 & 3 \\
 & Corrected hybrid & 0.5000 & 0.2887 & 0 & 4 \\
\bottomrule
\end{tabular}
\end{table}

In Q8.24 the published quadratic seed with one refinement reached 4.4184 LSB of maximum ideal error. A second refinement reduced this to 0.6133 LSB with two additional products. The degree-ten correction reached 0.7106 LSB with 11 products. The four-product corrected Hybrid and the exact radix-2 control returned the rounded reference for every input. The radix-2 control required 18 recurrence steps in Q16.16 and 26 in Q8.24. The two-step quadratic result improves maximum absolute error in Q16.16 but increases the number of rounded mismatches.

\begin{figure}[H]
\noindent The rounded-reference mismatch rates are shown in Figure~\ref{fig:baseline}.

\smallskip
  \centering
  \includegraphics[width=3.4in]{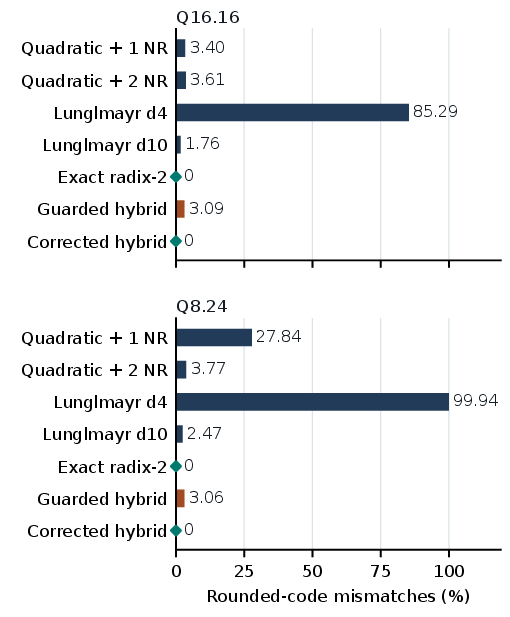}
  \Description{Horizontal bars for Q16.16 and Q8.24 compare rounded-code mismatch rates for quadratic refinement with one or two steps, the degree-four and degree-ten polynomial baselines, exact radix-2, guarded Hybrid and corrected Hybrid.}
  \caption{Rounded-reference mismatch rates for the polynomial baselines and the selected reciprocal alternatives.}
  \label{fig:baseline}
\end{figure}

\subsection{Portable RTL Results}\label{sec:hardwareresults}

Both portable correction methods and the exact radix-2 control returned the rounded reference on all 131,073 Q16.16 inputs and all 33,554,433 Q8.24 inputs. Every design sustained one accepted input per cycle under continuous valid input. The guarded Hybrid row and the full-product corrected row use four guard bits. The compact rows include matched and reduced guard widths and use Equation~(\ref{eq:reuse}).

With the guard width fixed at four bits, the compact datapath reduced the DSP count from six to five in Q16.16 and from 12 to 9 in Q8.24. The LUT and flip-flop counts increased in both formats. Reducing the guard widths lowered these logic counts relative to the four-guard-bit compact designs and preserved the DSP counts and the 10-cycle latency. The Q8.24 one-guard-bit variant uses eight fewer flip-flops than the two-guard-bit variant and six more LUTs. Its coefficient payload is smaller by 512 bits. The exact radix-2 alternative used no DSP slices and was deeper in both formats. These counts identify the tradeoff among logic, hard multipliers and pipeline depth for the tested RTL.

\begin{table}[H]
\noindent The exhaustive error, measured cycle counts and mapped resources are given in Table~\ref{tab:rtl}. Compact denotes the product-reuse corrected Hybrid.

\smallskip
\caption{Exhaustive RTL error, cycle latency and mapped resources for the same 7 series target family.}
\label{tab:rtl}
\centering
\small
\setlength{\tabcolsep}{2.6pt}
\begin{tabular}{@{}llrrrrrrr@{}}
\toprule
Format & Design & \shortstack[r]{Ideal max.\\(LSB)} & \shortstack[r]{Code max.\\(LSB)} & Cycles & LUTs & FFs & DSPs & \shortstack[r]{BRAM36\\equiv.}\\
\midrule
Q16.16 & Direct table & 123.883 & 124 & 1 & 7 & 14 & 0 & 2.0 \\
 & Guarded interpolation & 0.664 & 1 & 4 & 20 & 82 & 1 & 2.5 \\
 & Guarded hybrid & 0.597 & 1 & 8 & 86 & 185 & 4 & 0.0 \\
 & Corrected hybrid & 0.500 & 0 & 10 & 176 & 240 & 6 & 0.0 \\
 & Compact (\(g=4\)) & 0.500 & 0 & 10 & 231 & 282 & 5 & 0.0 \\
 & Compact (\(g=0\)) & 0.500 & 0 & 10 & 192 & 229 & 5 & 0.0 \\
 & Exact radix-2 & 0.500 & 0 & 20 & 688 & 762 & 0 & 0.0 \\
\addlinespace[2pt]
Q8.24 & Direct table & 32,732.039 & 32,732 & 1 & 11 & 14 & 0 & 3.0 \\
 & Guarded interpolation & 32.436 & 32 & 4 & 32 & 114 & 1 & 3.5 \\
 & Guarded hybrid & 0.646 & 1 & 8 & 288 & 286 & 8 & 0.0 \\
 & Corrected hybrid & 0.500 & 0 & 10 & 445 & 392 & 12 & 0.0 \\
 & Compact (\(g=4\)) & 0.500 & 0 & 10 & 496 & 422 & 9 & 0.0 \\
 & Compact (\(g=1\)) & 0.500 & 0 & 10 & 450 & 398 & 9 & 0.0 \\
 & Compact (\(g=2\)) & 0.500 & 0 & 10 & 444 & 406 & 9 & 0.0 \\
 & Exact radix-2 & 0.500 & 0 & 28 & 1,410 & 1,514 & 0 & 0.0 \\
\bottomrule
\end{tabular}
\end{table}

\subsection{Resource and Latency Tradeoffs}\label{sec:portable_tradeoffs}

\begin{table}[H]
\noindent Table~\ref{tab:portable_selection} gives examples of architecture selection after a numerical and resource budget has been imposed. The smallest simulated cycle count is selected among feasible designs. Equal cycle counts are resolved by the smaller DSP count and then the smaller mapped LUT count.

\smallskip
\caption{Portable designs selected under rounded-code and resource constraints. The error limit applies to the largest difference from the correctly rounded output over the entire tested input grid.}
\label{tab:portable_selection}
\centering
\small
\begin{tabular}{@{}>{\raggedright\arraybackslash}p{0.37\linewidth}
                 >{\raggedright\arraybackslash}p{0.28\linewidth}
                 >{\raggedright\arraybackslash}p{0.28\linewidth}@{}}
\toprule
Requirement & Q16.16 selection & Q8.24 selection \\
\midrule
Exact rounding & Compact ($g=0$), 10 cycles & Compact ($g=2$), 10 cycles \\
Exact rounding without DSP slices & Exact radix-2, 20 cycles & Exact radix-2, 28 cycles \\
At most one output code of error & Guarded interpolation, 4 cycles & Guarded Hybrid, 8 cycles \\
At most one output code of error without block RAM & Guarded Hybrid, 8 cycles & Guarded Hybrid, 8 cycles \\
\bottomrule
\end{tabular}
\end{table}

\begin{figure}[H]
\noindent Figure~\ref{fig:portable_resource_latency} compares the cycle latency and mapped logic of the portable implementations in Table~\ref{tab:rtl}. The compact corrected Hybrid and the full product corrector use the same number of cycles, but the compact variant uses fewer DSP slices at each matched format and guard width.

\smallskip
\centering
\includegraphics[width=\linewidth]{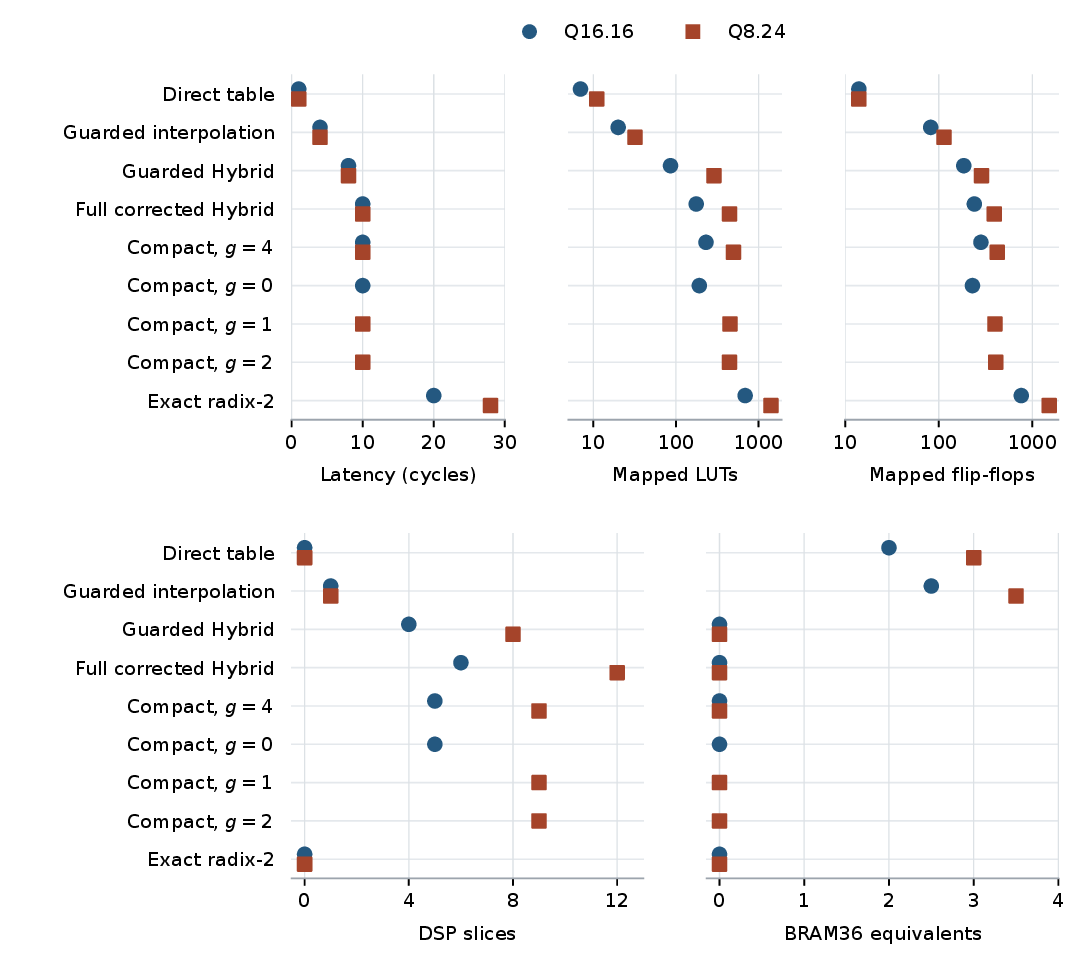}
\Description{Five horizontal dot plots show cycle latency, mapped lookup tables, flip-flops, DSP slices and block RAM for the two formats. Direct lookup has the shortest pipeline. Exact radix-2 has the deepest pipeline and uses no DSP slices or block RAM.}
\caption{Portable RTL latency and Yosys resource counts for Q16.16 and Q8.24. The lookup table and flip-flop axes are logarithmic.}
\label{fig:portable_resource_latency}
\end{figure}

The direct table needs the fewest logic LUTs and has the shortest pipeline, but its largest rounded-code error exceeds one LSB in both formats. The exact radix-2 pipeline avoids DSP slices by implementing its recurrence in logic and therefore has the largest LUT count. The compact corrected Hybrid uses DSP slices to reduce the number of dependent clock stages needed for exact rounding. These tradeoffs refer to the mapped cells and simulated cycle counts of the tested datapaths. A cycle count gives a time in seconds only after an operating clock period has been established for that implementation.

In Q16.16, the guarded interpolator is the shortest tested pipeline that stays within one rounded output code. It uses block RAM. If block RAM is unavailable, the guarded Hybrid meets the same code error requirement with a deeper pipeline. In Q8.24, guarded interpolation reaches a maximum error of 32 rounded output codes, so the one-code limit requires the guarded Hybrid among the two designs. When exact rounding is required, the compact Q8.24 variant with two guard bits is selected because it has six fewer mapped LUTs than the one-guard-bit variant. The latter uses eight fewer flip-flops and has a smaller coefficient payload, which may be preferable if either of those resources is more restrictive.

\section{Design Implications and Scope}\label{sec:discussion}

An application budget should identify whether it constrains relative error, absolute error or the exact rounded output. The allowed absolute error represented by a percentage depends on the input. The application may also limit table payload or the number of multiplications in the critical dependency chain. The exhaustive results support these choices on the stated input range (as shown in Sections~\ref{sec:numericalbudget} and~\ref{sec:comparison}). For a correctly rounded result and a DSP budget of zero, the unrolled radix-2 control is an option. Where DSP slices can be allocated, the compact Hybrid offers the same output codes with fewer pipeline cycles. A guarded interpolator remains useful when the application permits a larger absolute error and can allocate block memory. These choices are supported by the measured cycle and mapping data (as shown in Section~\ref{sec:hardwareresults}).

The analysis is confined to positive input values between 0.5 and 2.5. A larger input domain would require a normalization rule, output range checks and an analysis of its cost. The normalized baselines include their arithmetic shifts for the selected range. The original designs treat the range as a property of their input interface.

The reported resource counts describe technology mapping. An achieved clock rate requires routed timing evidence. Energy per result requires activity-based power evidence at a specified operating point. The difference between default switching assumptions and annotated activity is described in the Vivado power guide \cite{VivadoUG907}.

\section{Conclusion}\label{sec:conclusion}

Fixed-point reciprocal designs were evaluated for every Q16.16 and Q8.24 input code on the stated interval. Exact integer scoring exposed differences between approximation error, output quantization and correctly rounded code matches. A single integer residual adjustment was sufficient for exact rounding in both tested formats. The compact implementation passed exhaustive RTL verification and preserved one accepted input per cycle. With a latency of 10 cycles, its Q8.24 implementation mapped to nine DSP slices. The exact radix-2 alternative mapped to no DSP slices and required 28 cycles. Resource and throughput conclusions are limited to the implementation flow used for the corresponding RTL. The exhaustive error budgets show where extra fractional bits improve accuracy and where the address width limits it. The supplied evidence allows numerical and mapped hardware tradeoffs to be recalculated for other constraints on the same range.

\bibliographystyle{ACM-Reference-Format}
\bibliography{fpga-reciprocal-references}

@manual{AMD_PG151,
  author = {{Advanced Micro Devices}},
  title = {Divider Generator v5.1 Product Guide},
  organization = {AMD},
  number = {PG151},
  year = {2021},
  url = {https://docs.amd.com/v/u/en-US/pg151-div-gen}
}

@book{ErcegovacLang04,
  author = {Ercegovac, Milos D. and Lang, Tomas},
  title = {Digital Arithmetic},
  publisher = {Morgan Kaufmann},
  address = {San Francisco, CA, USA},
  year = {2004}
}

@article{EvenGoldschmidt05,
  author = {Even, Guy and Seidel, Peter-Michael and Ferguson, Warren E.},
  title = {A Parametric Error Analysis of {Goldschmidt}'s Division Algorithm},
  journal = {Journal of Computer and System Sciences},
  year = {2005},
  volume = {70},
  number = {1},
  pages = {118--139},
  doi = {10.1016/j.jcss.2004.08.004},
  url = {https://doi.org/10.1016/j.jcss.2004.08.004}
}

@mastersthesis{Goldschmidt64,
  author = {Goldschmidt, Robert E.},
  title = {Applications of Division by Convergence},
  school = {Massachusetts Institute of Technology},
  address = {Cambridge, MA, USA},
  year = {1964},
  type = {M.S. thesis},
  url = {https://dspace.mit.edu/entities/publication/790ae046-6e77-4fb2-8cb6-618e3b86647f}
}

@article{Gutierrez26,
  author = {Guti{\'e}rrez-Ramos, Gonzalo and Parra-Michel, Ram{\'o}n and Romero-Aguirre, Eduardo and Rodriguez-Garc{\'i}a, Alberto and Jaramillo-Ram{\'i}rez, Rodrigo},
  title = {A Fast Fixed-Point Implementation for Division, Reciprocal, Square Root and Reciprocal Square Root Based on Newton-Raphson Method},
  journal = {Electronics},
  volume = {15},
  number = {13},
  articleno = {2899},
  note = {Art. no. 2899},
  numpages = {18},
  year = {2026},
  doi = {10.3390/electronics15132899},
  url = {https://doi.org/10.3390/electronics15132899}
}

@article{Ito97,
  author = {Ito, M. and Takagi, N. and Yajima, S.},
  title = {Efficient Initial Approximation for Multiplicative Division and Square Root by a Multiplication With Operand Modification},
  journal = {IEEE Transactions on Computers},
  volume = {46},
  number = {4},
  pages = {495--498},
  year = {1997},
  month = apr,
  doi = {10.1109/12.588066},
  url = {https://doi.org/10.1109/12.588066}
}

@article{Kumar17,
  author = {Kumar, D. and Saha, P. and Dandapat, A.},
  title = {Hardware Implementation of Methodologies of Fixed Point Division Algorithms},
  journal = {International Journal on Smart Sensing and Intelligent Systems},
  volume = {10},
  number = {3},
  pages = {1--16},
  year = {2017},
  doi = {10.21307/ijssis-2017-227},
  url = {https://doi.org/10.21307/ijssis-2017-227}
}

@misc{KuttyArtifact,
  author = {Kutty, Navaneeth},
  title = {Fixed-Point Research},
  year = {2026},
  month = oct,
  note = {Revision 49350ee, November 14, 2026},
  url = {https://github.com/navaneethkutty/fixed-point-research/tree/49350ee}
}

@article{Libessart17,
  author = {Libessart, E. and Arzel, M. and Lahuec, C. and Andriulli, F. P.},
  title = {A Scaling-Less Newton-Raphson Pipelined Implementation for a Fixed-Point Reciprocal Operator},
  journal = {IEEE Signal Processing Letters},
  volume = {24},
  number = {6},
  pages = {789--793},
  year = {2017},
  month = jun,
  doi = {10.1109/LSP.2017.2694225},
  url = {https://doi.org/10.1109/LSP.2017.2694225}
}

@misc{Lunglmayr21,
  author = {Lunglmayr, Michael},
  title = {Efficient Non-sequential Division for {FPGAs}},
  year = {2022},
  eprint = {2105.05747v2},
  note = {Version 2, September 9, 2022},
  url = {https://arxiv.org/abs/2105.05747v2},
  archivePrefix = {arXiv},
  doi = {10.48550/arXiv.2105.05747}
}

@article{Oberman97,
  author = {Oberman, Stuart F. and Flynn, Michael J.},
  title = {Division Algorithms and Implementations},
  journal = {IEEE Transactions on Computers},
  volume = {46},
  number = {8},
  pages = {833--854},
  year = {1997},
  month = aug,
  doi = {10.1109/12.609274},
  url = {https://doi.org/10.1109/12.609274}
}

@inproceedings{Rodriguez13,
  author = {Rodr{\'i}guez-Garc{\'i}a, A. and Pizano-Escalante, L. and Parra-Michel, R. and Longoria-Gandara, O. and Cortez, J.},
  title = {Fast Fixed-Point Divider Based on Newton-Raphson Method and Piecewise Polynomial Approximation},
  booktitle = {Proceedings of the International Conference on Reconfigurable Computing and FPGAs (ReConFig)},
  pages = {1--6},
  year = {2013},
  publisher = {IEEE},
  address = {Piscataway, NJ, USA},
  location = {Cancun, Mexico},
  doi = {10.1109/ReConFig.2013.6732291},
  url = {https://doi.org/10.1109/ReConFig.2013.6732291}
}

@manual{VerilatorDocs,
  author = {{Verilator Project}},
  title = {Verilator User's Guide},
  year = {2026},
  url = {https://verilator.org/guide/latest/},
  note = {Accessed November 11, 2026}
}

@manual{VivadoUG907,
  author = {{Advanced Micro Devices}},
  title = {Vivado Design Suite User Guide for Power Analysis and Optimization},
  organization = {AMD},
  number = {UG907},
  year = {2025},
  note = {Version 2025.2},
  url = {https://docs.amd.com/r/2025.2-English/ug907-vivado-power-analysis-optimization/Supported-Inputs}
}

@manual{XilinxUG479,
  author = {{Xilinx, Inc.}},
  title = {7 Series DSP48E1 Slice User Guide},
  organization = {Xilinx, Inc.},
  number = {UG479},
  edition = {1.10},
  year = {2018},
  month = mar,
  url = {https://docs.amd.com/v/u/en-US/ug479_7Series_DSP48E1}
}

@manual{YosysXilinx,
  author = {{YosysHQ}},
  title = {Yosys Documentation for Xilinx FPGA Synthesis},
  year = {2025},
  note = {Yosys 0.52, Xilinx synthesis command},
  url = {https://yosyshq.readthedocs.io/projects/yosys/en/v0.52/cmd/synth_xilinx.html}
}

\appendix
\section{Reproduction and Data Availability}\label{sec:reproduce}

The original RTL revision is identified by its Git commit \cite{KuttyArtifact}. The accompanying simulation archive contains the independent numerical and RTL code, parameter settings, result tables and figure generation commands. The numerical evaluation uses a C++20 compiler. Python with Matplotlib is used to generate the figures. The input codes at the reported maxima are recorded in the result files. The corresponding outputs can be regenerated with the supplied models. The reproducibility archive includes the signed-error histograms, percentile and input-bin data, numerical budget selections, Pareto comparisons and the synthesis records behind the portable design selections. Its build instructions describe how to generate the RTL evidence and the Encapsulated PostScript (EPS) figures. The dataset contains no records from external collections.

\section{Supplementary Error Distributions}\label{app:error_distributions}

The histogram in this appendix uses $\log_{10}(1+|q-q^\star|)$ to keep exact matches at zero and make larger rounded-code differences visible. Each bin has width 0.15 on the transformed scale. The conditional quartiles in Q8.24 use only the inputs whose outputs differ from the rounded reference.

\begin{figure}[H]
\noindent The transformed histograms and conditional quartiles are shown in Figure~\ref{fig:supplementary_magnitudes}.

\smallskip
  \centering
  \includegraphics[width=\linewidth]{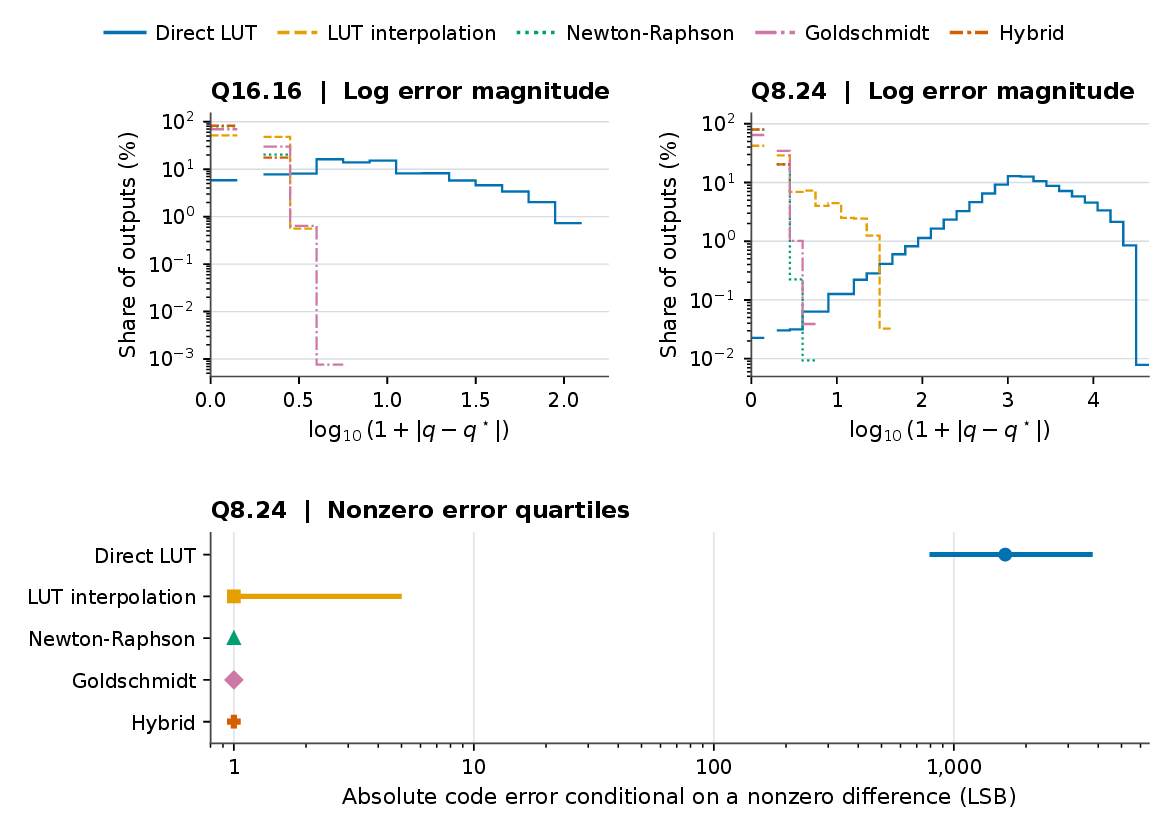}
  \Description{The two upper panels show histogram shares of the logarithm of one plus the absolute rounded-code difference. The lower panel shows Q8.24 medians and interquartile ranges among nonzero differences. Direct lookup has the largest interquartile range.}
  \caption{Logarithmic absolute code error histograms on both exhaustive grids and Q8.24 nonzero error medians with interquartile ranges.}
  \label{fig:supplementary_magnitudes}
\end{figure}

Among the nonzero Q8.24 differences, the lower quartile, median and upper quartile are 793, 1{,}637 and 3{,}786 LSB for direct lookup. They are one, one and five LSB for interpolation and one LSB at all three quartiles for each refined path. These conditional distributions describe the size of a mismatch when one occurs. The full-grid agreement rates are reported in Table~\ref{tab:cross_format}.

\end{document}